\documentclass[conference,10pt]{IEEEtran}
\usepackage{amsmath}
\usepackage{amsfonts}
\usepackage{amssymb}
\usepackage{setspace}
\usepackage[latin1]{inputenc}
\usepackage{graphicx}
\usepackage{amsthm}
\usepackage{color}
\usepackage{cite}
\usepackage{bm}
\usepackage{epsfig,psfrag}
\usepackage{tabularx}
\usepackage{tikz}
\usepackage{pst-node}
\usepackage{acronym}
\usepackage[yyyymmdd,hhmmss]{datetime}
\usepackage{notation}
\usetikzlibrary{arrows,backgrounds,calc,positioning,shapes,shadows}
\usepackage{gensymb}

\newcommand{\ist}{\hspace*{.3mm}}
\newcommand{\rmv}{\hspace*{-.3mm}}

\newcommand{\nn}{\nonumber}

\newcommand{\trans}{^\mathrm{T}}

\definecolor{greenNew}{RGB}{0,170,0}
\definecolor{grayNew}{RGB}{50,50,50}

\newcommand{\rd}{\textcolor{red}}
\newcommand{\bl}{\textcolor{blue}}
\newcommand{\gre}{\textcolor{greenNew}}
\newcommand{\gra}{\textcolor{grayNew}}

\allowdisplaybreaks

\begin{document}
\title{Acoustic Source Localization in Shallow Water: \\[-.5mm] A Probabilistic Focalization\vspace*{-1mm} Approach}

\author{\IEEEauthorblockN{Florian Meyer\IEEEauthorrefmark{1} and 
Kay L. Gemba\IEEEauthorrefmark{2}}\vspace{.8mm}
\IEEEauthorblockA{\IEEEauthorrefmark{1}Scripps Institution of Oceanography and Department of Electrical and Computer Engineering\\ University of California San Diego, La Jolla, CA (flmeyer@ucsd.edu)\vspace{.8mm}}
\IEEEauthorblockA{\IEEEauthorrefmark{2}Physics Department, Naval Postgraduate School, Monterey, California 93943, USA (kgemba@nps.edu)}
 \vspace*{-7mm}}

\maketitle

\begin{abstract}
This paper presents a Bayesian estimation method for the passive localization of an acoustic source in shallow water. Our probabilistic focalization approach estimates the time-varying source location by associating direction of arrival (DOA) observations to DOAs predicted based on a statistical model. Embedded ray tracing makes it possible to incorporate environmental parameters and characterize the nonlinear acoustic waveguide. We demonstrate performance advantages of our approach compared to matched field processing using data collected during the  \emph{SWellEx-96}  experiment. \vspace{3mm}\end{abstract}


\acresetall
\section{Introduction} \label{sec:intro}

In the shallow water environment, knowledge of the acoustic waveguide can be used to exploit multipath propagation for source localization. In particular, it has been demonstrated that a typical shallow water waveguide offers sufficient coherence times and ray-path diversity (or aperture) to determine the range and depth of an acoustic source from the measurements provided by a single vertical line array (VLA)\cite{KupHodHeeAkaFerJac:J98}.
\vspace{.5mm}

\subsection{State-of-the-Art}
Matched field processing\cite{Hin:J73,Buc:J76,BagKupMik:J93} (MFP) is a scientific approach to shallow water localization that compares modeled signal replicas with acoustic data in order to determine the location of one or multiple sources. A potential limitation of MFP is the fact that modeling the channel between a candidate source position and a VLA requires detailed knowledge of environmental parameters.
\begin{figure*}[t!]
\centering
\hspace{5mm}\begin{minipage}[H!]{0.35\textwidth}
\centering
\vspace{-3mm}
\includegraphics[scale=.45]{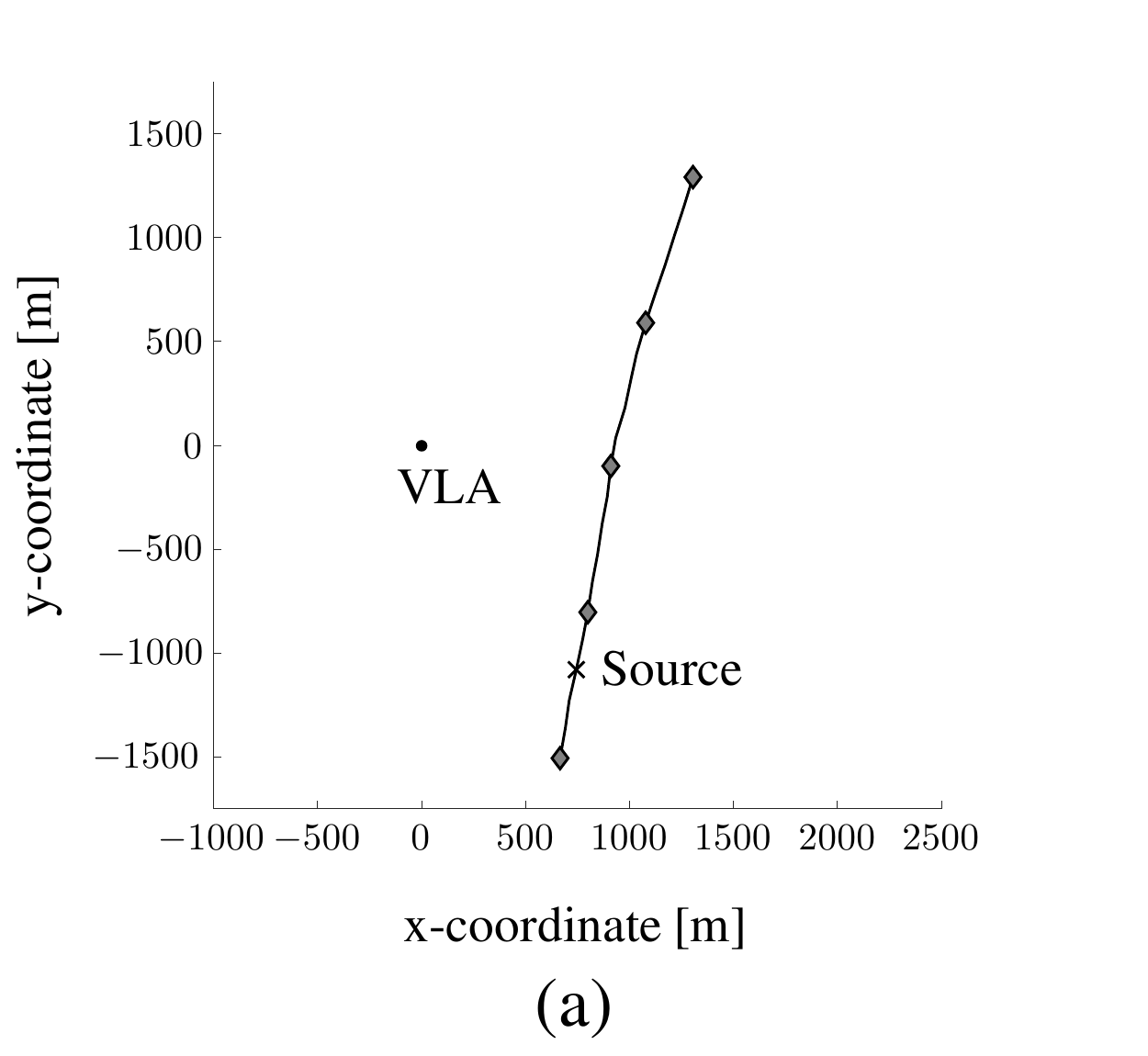}
\vspace{4mm}
\end{minipage}
\hspace{0mm}\begin{minipage}[H!]{0.60\textwidth}
\centering
\vspace{2.8mm}
\includegraphics[scale=.45]{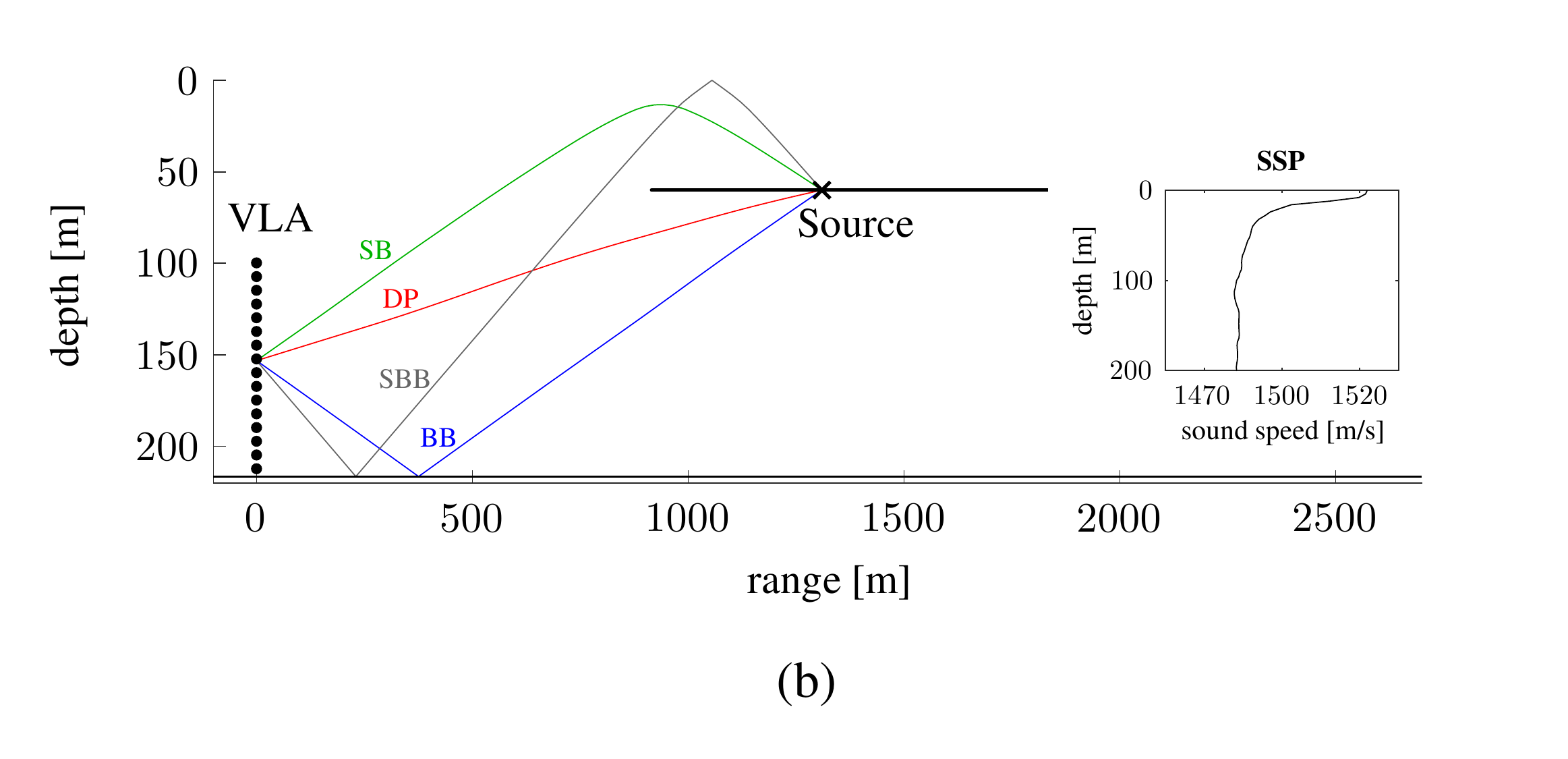}
\vspace{4mm}
\end{minipage}
\renewcommand{\baselinestretch}{1.05}\small\normalsize
\vspace{-3mm}
\caption{Considered shallow water localization scenario from the \emph{SWellEx-96} experiment with VLA position and source track in the time interval 11-May-1996 00:05:00--00:25:00 UTC. (a) shows the track of the acoustic source towed by the R/V Sproul at a depth of roughly 60m. The black cross shows the source position at time 00:08:00~UTC and gray diamonds indicate source positions in 5 minutes intervals. (b) shows the VLA and source track in range and depth. The black cross indicates the same source position as in (a). The $K = 4$ dominant propagation paths \rd{direct path (DP)}, \gre{surface bounce (SB)}, \bl{bottom bounce (BB)}, and \gra{surface-bottom bounce (SBB)}, are shown.}
\label{fig:rayTracing}
\vspace{1mm}
\end{figure*}

MFP is still an active research topic\cite{ColFiaKup:J95,ThoKupDsp:J00,ByuVerSab:J17,GemNanGerHod:J17,GemHodGer:J17,OrrNicPer:J20} because in real-world scenarios, knowledge of the acoustic environment, in particular an accurate model of the seabed, is typically unavailable. Recent approaches to shallow water localization explicitly consider environmental uncertainty to offer a degree of robustness to model mismatch\cite{Kro:J92,GemNanGerHod:J17,ByuHunGemSonKup:J20}.

\subsection{Contribution and Paper Organization}

This paper presents an innovative approach for the localization and tracking of an underwater source with a probabilistic model of the environment.  Signals transmitted by an acoustic source are received by a VLA. Inspired by recently proposed graph-based methods for multitarget tracking and indoor localization and mapping\cite{MeyBraWilHla:J17,MeyKroWilLauHlaBraWin:J18,LeiMeyHlaWitTufWin:J19,MenMeyBauWin:J19,MeyWin:J20,MeyWil:J21}, the proposed sequential probabilistic focalization\cite{Col:J91} method associates direction of arrival (DOA) observations to modeled DOAs and jointly estimates the time-varying position of the source. Embedded ray tracing makes it possible to incorporate environmental parameters such as the SSP (sound speed profile) of the acoustic channel. Evaluation of the proposed method is performed based on acoustic data from the \emph{SWellEx-96} shallow water localization scenario shown in Fig.~\ref{fig:rayTracing}. The contributions of this paper are as follows.
\begin{itemize}
\item We introduce a new probabilistic model for the localization of an acoustic source in the shallow water waveguide by using a\vspace{1mm}  VLA.
\item We establish a Bayesian estimation method based on the new model and evaluate its performance using data collected during the  \emph{SWellEx-96}  experiment.
\end{itemize}

An important aspect  is the \emph{probabilistic data association} of DOA observations to modeled propagation paths for a source and VLA geometry. The considered model calculates expected DOAs by means of ray tracing based on the SSP and, if available, a characterization of the\vspace{1mm} seabed. 

\section{System Model} \label{sec:sysModel}

We consider a mobile source with unknown time-varying position $\V{p}_n \rmv\!\in\rmv \mathbb{R}^2$ which, due to the azimuthal ambiguity of DOA information provided by the VLA, only consists of range and depth. There are $K$ propagation paths that are used for localization. In each discrete time slot $n$, the array acts as a receiver and provides DOA observations.  A DOA estimation method \cite{NanGemGerHodMec:J19,MeyParGer:C20} processes the acoustic signals received by the VLA and returns $M_n$ DOA estimates at time $n$. $M_n$ at time $n$ is related to the number of propagation paths $K$ as follows: It is possible that some propagation paths are not ``detected'' by the DOA estimator and thus do not generate a DOA estimate, and it is also possible that some DOA estimates do not correspond to a propagation path. Accordingly, $M_n$ may be smaller than, equal to, or larger than $K$.  Note also that $M_n$ depends on the source position $\V{p}_n$ and on the environment. Selecting the number of propagation paths $K$ requires some a priori understanding of the propagation environment. 

\begin{figure*}[t!]
\centering
\vspace{-1mm}
\hspace*{-3.5mm}\includegraphics[scale=.8]{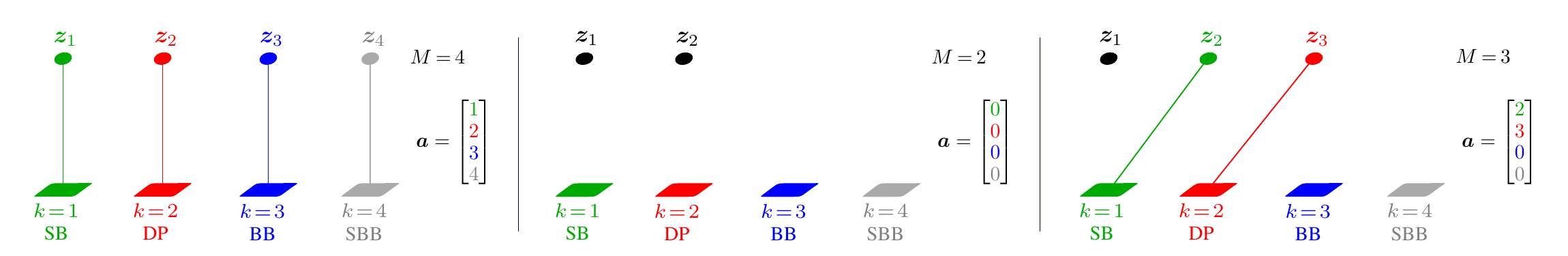}
\vspace{-5mm}
\caption{Three example realizations of DOA observations $\V{z}_n \rmv\triangleq\rmv \big[z_{1,n}\trans\ist \cdots\ist z_{M_n,n}\trans \big]\trans\rmv\vspace{.5mm}$ and data association vector $\V{a}_n$ for the scenario Fig.~\ref{fig:rayTracing}. Associations of DOA observations to the $K \rmv=\rmv4$ modeled propagation paths \gre{surface bounce (SB)}, \rd{direct path (DP)},  \bl{bottom bounce (BB)}, and \gra{surface-bottom bounce (SBB)}, are shown. The time index $n$ is omitted.}
\vspace{-1mm}
\label{fig:daEvents}
\end{figure*}

\subsection{Source State and Association Vectors}
\label{sec:vec_description}

The state of the source at time $n$ is $\V{x}_n \!\triangleq [\V{p}_n\trans \; v_n]\trans\rmv$, where $v_n$ is the source speed in range. The source state $\V{x}_n$ is assumed to evolve according to Markovian state dynamics, where $f(\V{x}_{n}|\V{x}_{n-1})$ is the state-transition probability density function (PDF) of the source state. At time $n\rmv= 0$, the source state $\V{x}_0$ is distributed according to a uninformative prior PDF $f(\V{x}_0)$.

The DOA observations $z_{m,n} \rmv\in\rmv [-90^{\circ},90^{\circ})$, $m \rmv\in\{1,\dots,M_n\}$ are subject to observation-origin uncertainty. That is, it is not known which observation $z_{m,n}\ist$ is associated with which propagation path $\{1,\dots,K\}$, or if an observation $z_{m,n}$ did not originate from any propagation path (this is known as a \emph{false alarm}), or if a propagation path did not give rise to any observation (this is known as a \emph{missed detection}). The probability that a propagation path is ``detected'' in the sense that it generates an observation $z_{m,n}$ in the DOA estimation stage is denoted by $d_k\big(\V{p}_n \big) \!\in\rmv [0,1]$. (For positions $\V{p}_n$ for which a propagation path $k$ is geometrically impossible, we set $d_k\big(\V{p}_n \big) = 0$.) False alarms are independent and identically distributed as $f_{\text{FA}}\big( z_{m,n} \big)$. The number of false alarms is assumed Poisson distributed with mean $\mu_\text{FA}\rmv$ \cite{BarWilTia:B11,MeyBraWilHla:J17,MeyKroWilLauHlaBraWin:J18}.  $f_{\text{FA}}\big( z_{m,n} \big)$, and $\mu_\text{FA}\rmv$ are known. The DOA observations resulting from propagation paths that are not modeled among the $K$ selected paths are treated as false alarms. Finally, we introduce the total observation vector $\V{z}_n \rmv\triangleq\rmv \big[z_{1,n}\ist \cdots\ist z_{M_n,n} \big]\trans\rmv$ that is assumed sorted in descending order, i.e., $z_{m,n} \rmv\geq\rmv z_{m+1,n}$, $m \rmv\in\{1,\dots,M_n\rmv-\rmv1\}$.

The associations between DOA observation $m \in \{1,\dots,M_n\}$ and propagation path $k \in \{1,\dots,K\}$ at time $n$ can be described by the $K$-dimensional \vspace{0mm} \textit{data association vector} $\V{a}_n = \big[a_{1,n} \cdots\ist a_{K,n} \big]\trans\rmv\rmv$, whose $k$th entry is given by $a_{k,n} \!=\rmv m \!\in\! \{1,\dots,M_n\}$ if propagation path $k$ generates observation $z_{m,n}$, and $a_{k,n} \!=\rmv 0$ if  it does not generate any observation. Due to observation-origin uncertainty, $\V{a}_n$ is a random variable. However, since the DOAs of the propagation paths have a fixed order\cite{FinKupPor:B11} and each propagation path can generate at most one DOA observation\cite{BarWilTia:B11,MeyBraWilHla:J17,MeyKroWilLauHlaBraWin:J18} at any time $n$, only certain data association vectors $\V{a}_n$ are valid. To facilitate identifying invalid vectors $\V{a}_n$, we order its elements $a_{1,n},\dots,a_{K,n}$ such that for the case where there are no false alarms and no \vspace{.3mm} missed detections, we \vspace{.4mm} have $\V{z}_n \rmv= \big[z_{a_{1,n},n}\ist \cdots\ist z_{a_{K,n},n} \big]\trans\rmv\rmv\rmv$. With this order a general association vector $\V{a}_n$ is invalid if and only if there exist $k,k' \!\in \{1,\dots,K\}$ with $k \rmv>\rmv k'\rmv$ such that $a_{k',n} \!\geq\rmv a_{k\rmv,n} \!\neq\rmv 0$.

For examples, let \vspace{.5mm} us assume that we use the $K \rmv=\rmv 4$ propagation paths shown in Fig.~\ref{fig:rayTracing}. According to the order defined above, we have  $k = 1$ for the SB, $k = 2$ for the DP, $k = 3$ for the BB, and $k\rmv=\rmv4$ for the SBB. It can easily be verified that for the case where there is a measurement for each propagation path and no false alarm, we obtain $a_{1,n} \rmv=\rmv 1$, $a_{2,n} \rmv=\rmv 2$, $a_{3,n} \rmv=\rmv 3$, and $a_{4,n} \rmv=\rmv 4$ as well as $z_{1,n} \geq z_{2,n} \geq z_{3,n} \geq z_{4,n}$ (cf. with the DOAs of the four propagation paths at the VLA shown in Fig.~\ref{fig:rayTracing}).

\subsection{Observation Model}
\label{sec:meas_likelihood} 

At each time step $n$, a DOA estimation method \cite{NanGemGerHodMec:J19,MeyParGer:C20} provides DOAs $z_{m,n}$, $m \rmv\in\rmv \{1,\dots,M_n\}$. These DOAs are considered as the observations in our statistical model. Let $z_{m,n}$ be the DOA observations related to propagation path $k \rmv\in\rmv \{1,\dots,K\}$. The observation model is then given \vspace{0mm} by
\begin{equation}
z_{m,n} = g_k(\V{p}_n) + w_{k,n}.
\vspace{1.5mm}
\label{eq:measModel}
\end{equation}
Here, $w_{k,n}$ is zero mean Gaussian distributed with variance $\sigma^2_{k}$ and $g_k(\V{p}_n)$ is the DOA of propagation path $k$. The observation noise $w_{k,n}$ assumed statistical independent across time $n$ and propagation path $k$. From \eqref{eq:measModel}, we directly get the conditional PDF $f_k(z_{m,n}|\V{x}_{n})$. The functions $g_k(\V{p}_n)$ are the DOAs of eigenrays related to the considered propagation paths. The eigenrays are obtained using the BELLHOP\cite{Por:I20} ray-tracing software that can incorporate the SSP and a model of the seabed. Details on how the $g_k(\V{p}_n), k \in \{1,\dots,K\}$ can be precomputed offline will be discussed in Section~\ref{sec:impAspects}.

\subsection{Measurement Model}
\label{sec:joint_likelihood} 

The conditional PDFs $f_k\big(z_{m,n} \big| \V{x}_n \big)$ obtained from \eqref{eq:measModel} characterize the statistical relation between the observations $z_{m,n}$ and the states $\V{x}_n$.  This PDF is a central element\vspace{-.4mm} in the conditional PDF of the total observation vector $\V{z}_n$ given $\V{x}_n$, $\V{a}_n$, and $M_n$. Let us introduce the set of detected paths at time $n$ as  $\Set{D}_{\V{a}_n} \rmv\triangleq\rmv \big\{ k \in \{1,\dots,K\} \ist \big| \ist a_{k,n} \neq 0\big\}$. Conditioned on $\V{x}_n$, we assume that DOAs generated by the source are statistical independent of false alarms, i.e., we can write
\begin{align}
&f\big( \V{z}_{n} \big| \V{x}_{n}, \V{a}_{n} \rmv, M_{n}\big) \nn\\
&\hspace{2mm} = \Bigg( \prod^{M_n}_{m = 1} f_{\text{FA}}\big( z_{m,n} \big) \rmv\Bigg) \!\left(\ist\prod_{k\in \Set{D}_{\V{a}_n}}\!\! \frac{f_k\big( z_{a_{k,n},n} \big|\ist \V{x}_n \big)} {f_{\text{FA}}\big( z_{a_{k,n},n} \big)}\ist\right)
\label{eq:condPDF}\\[-5mm]
\nn
\end{align}
if the number of elements in the vector $\V{z}_{n}$ is equal to $M_{n}$ and $f\big( \V{z}_{n} \big| \V{x}_{n}, \V{a}_{n} \rmv, M_{n}\big) \rmv=\rmv 0$, otherwise.
For example, let us again assume $K = 4$ and discuss three example realizations of $M_n$ and $\V{a}_n$. These three example realizations are shown in in Fig.~\ref{fig:daEvents}. In the ideal case where there is no missed detection and no false alarm, i.e., $M_n = 4$ and $\V{a}_n = [1 \ist \ist 2 \ist \ist  3 \ist \ist  4]^{\trans} \rmv\rmv\rmv\rmv$, the conditional PDF in \eqref{eq:condPDF} reads $f\big( \V{z}_{n} \big| \V{x}_{n}, \V{a}_{n} \rmv, M_{n}\big)  = \prod^4_{k=1} f_k\big( z_{k,n} \big| \V{x}_n \big)$.  Similarly, in case there is no detected path and two false alarms, i.e., $M_n = 2$ and $\V{a}_n = [0 \ist \ist 0 \ist \ist  0 \ist \ist  0]^{\trans}\rmv\rmv$, we have $f\big( \V{z}_{n} \big| \V{x}_{n}, \V{a}_{n} \rmv, M_{n}\big)  = \prod^2_{m=1} f_{\text{FA}}\big( z_{m,n} \big) $. Finally, if we only detect the DP and the SB and there is also one missed detection, e.g., $M_n = 3$ and $\V{a}_n = [2 \ist \ist 3\ist \ist  0 \ist\ist 0\ist \ist]^{\trans}\rmv\rmv$, we have $f\big( \V{z}_{n} \big| \V{x}_{n}, \V{a}_{n} \rmv, M_{n}\big)  = f_{\text{FA}}\big( z_{1,n} \big)$ $f_1\big( z_{2,n} \big| \V{x}_n \big) \ist f_2\big( z_{3,n} \big| \V{x}_n \big)$.

Let us consider $f\big( \V{z}_{n} \big| \V{x}_{n}, \V{a}_{n} \rmv, M_{n}\big)$ as a \emph{likelihood function}, i.e., a function of $\V{x}_n$, $\V{a}_n\rmv$, and $M_{n}$, for observed $\V{z}_{n}\rmv$. If $\V{z}_{n}$ is observed and therefore fixed, also $M_{n}$ is fixed, and we can rewrite \eqref{eq:condPDF}, 
up to a constant normalization factor, \vspace{-1mm} as
\begin{align}
&f\big( \V{z}_n\big|\V{x}_n, \V{a}_n\rmv, M_{n}\big) \propto \ist \prod^{K}_{k=1} \ist h_k\big( \V{x}_{n}, a_{k,n}; \V{z}_{n} \big). 
\nn \\[-6mm]
\nn
\end{align}
Here, the factors $h_k\big( \V{x}_{n}, a_{k,n}; \V{z}_{n} \big)$ are \vspace{.8mm} defined as
\begin{align}
 h_k\big( \V{x}_{n}, a_{k,n}; \V{z}_{n} \big) &\triangleq \begin{cases}
  \rmv\displaystyle \frac{f_k\big(z_{a_{k,n},n} \big|\V{x}_n\big)}{f_{\text{FA}}\big( z_{a_{k,n},n} \big)} \ist, 
  &\!\rmv\rmv\rmv a_{k,n} \!\rmv\in\! \{1,\dots,M_n\} \\[.5mm]
  \rmv 1 \ist, & \!\rmv\rmv\rmv a_{k,n} \!=\rmv 0.
\end{cases} \nn
\end{align}
Finally, the likelihood function for $\V{z}_{1:n} \!\triangleq\rmv \big[\V{z}_1\trans \cdots\ist \V{z}_n\trans \big]\trans\rmv\rmv$, involving the observations $z_{m,n'}$ of  all time steps $n' \!=\! 1,\ldots,n$, is obtained \vspace{-1mm} as
\begin{align}
&\hspace{-1mm}f( \V{z}_{1:n} \ist|\ist \V{x}_{1:n},\V{a}_{1:n},\rmv\V{m}_{1:n}) = \ist \prod^{n}_{n'=1} \ist f\big( \V{z}_{n'}\big|\V{x}_{n'}, \V{a}_{n'}\rmv, M_{n'}\big)
\label{eq:allTimeslhf} \\[-8mm]
\nn
\end{align}
where we introduced $\V{m}_n = \big[M_1 \cdots M_n\big]^{\trans}$. (Recall that $M_n$ is the number of detected DOAs at time $n$.)
\vspace{-.8mm} 

\begin{figure}[t!]
\centering
\vspace{-1mm}
\hspace*{-3.5mm}\includegraphics[scale=1.2]{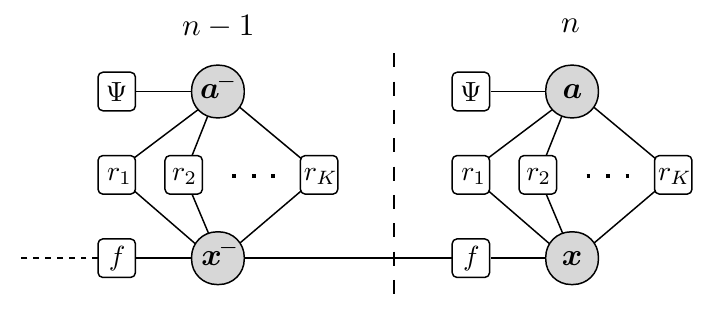}
\vspace{-5mm}
\caption{Factor graph representing the factorization of the joint posterior PDF $f(\V{x}_{1:n}, \V{a}_{1:n} | \V{z}_{1:n} )$ in \eqref{eq:factorOverall} depicted for two time steps $n' \rmv\in\rmv \{n \rmv-\rmv1,n\}$. The time index is omitted and the following short notations are used: $f \triangleq f(\V{x}_{n'} | \V{x}_{n'-1})$, $r_k \triangleq r_k\big( \V{x}_{n'}, a_{k,n'}; \V{z}_{n'}  \big) $, $\V{x} \triangleq \V{x}_n$, $\V{a} \triangleq \V{a}_n$, $\V{x}^{-} \triangleq \V{x}_{n-1}$, and $\V{a}^{-} \triangleq \V{a}_{n-1}$.}
\vspace{-4.5mm}
\label{fig:overallFG}
\end{figure}

\subsection{Prior Information}
\label{eq:JointAssociation}

Now, one can derive \cite{BarWilTia:B11,MeyKroWilLauHlaBraWin:J18} the following expression of the joint conditional prior probability mass function (pmf) of the data association vector $\V{a}_{n}$, and the number of observations $M_n$, conditioned on the state $\V{x}_n$ of the mobile \vspace{1mm} source, 
\begin{align}
p\big(\V{a}_{n}, M_{n} \big|\V{x}_n \big) &=\ist  \frac{e^{-\mu_\text{FA}} \ist \mu_\text{FA}^{M_{n} - |\Set{D}_{\!\V{a}_{n}}|} \ist\ist |\Set{D}_{\!\V{a}_{n}}| !}{M_{n}! }
  \ist \bigg( \prod_{k \in \Set{D}_{\V{a}_n}} \!\! d_{k}(\V{p}_n) \! \bigg)  \nn\\[2.5mm]
&\hspace{8mm} \times \psi\big(\V{a}_{n}\big)  \!\prod_{k'\rmv \notin \Set{D}_{\V{a}_n}} \!\!\!\!\big(1- d_{k'}(\V{p}_n) \big) \ist.
\label{eq:assocprior}\\[-6mm]
\nn
\end{align}
Here, $\psi\big(\V{a}_{n}\big)$ checks the validity of association vector $\V{a}_{n}$ as discussed in Section \ref{sec:vec_description}. In particular, it is defined to be $0$ if there exist $k,k' \!\in \{1,\dots,K\}$ with $k \rmv>\rmv k'\rmv$ such that $a_{k',n} \!\geq\rmv a_{k\rmv,n} \!\neq\rmv 0$,
and to be $1$ otherwise. The function $\psi\big(\V{a}_{n}\big)$ enforces $p\big(\V{a}_{n}, M_{n} \big|\V{x}_n \big) \rmv=\rmv 0$ if any observation is associated with more than one propagation path. For future reference, we can also express \eqref{eq:assocprior} as
\begin{align}
p\big(\V{a}_{n}, M_{n} \big|\V{x}_n \big)&\propto\ist  C(M_n) \ist\ist \psi\big(\V{a}_{n}\big) \ist \prod^{K}_{k=1} \ist v\big( \V{x}_{n}, a_{k,n}; M_{n} \big) \nn\\[-5.5mm]
\nn
\end{align}
 where the factors $v\big( \V{x}_{n}, a_{k,n}; M_{n} \big)$ are \vspace{.8mm} defined as
\begin{align}
 v\big( \V{x}_{n}, a_{k,n}; M_{n} \big) &\triangleq \begin{cases}
 \frac{d_{k}(\V{p}_n)}{\mu_{\text{FA}}}
  &\!\rmv\rmv\rmv a_{k,n} \!\rmv\in\! \{1,\dots,M_n\} \\[.5mm]
  \rmv 1- d_{k}(\V{p}_n) \ist, & \!\rmv\rmv\rmv a_{k,n} \!=\rmv 0.
\end{cases} \nn
\end{align}
and $C(M_n) \rmv=\rmv e^{-\mu_\text{FA}} \ist \mu_\text{FA}^{M_{n}} |\Set{D}_{\!\V{a}_{n}}| ! /M_{n}!$.

Using common assumptions\cite{BarWilTia:B11,MeyBraWilHla:J17,MeyKroWilLauHlaBraWin:J18}, the joint prior distribution for all source states, association variables, and number of observations up to time $n$, is given by
\begin{align}
&f(\V{x}_{1:n},\V{a}_{1:n}, \V{m}_{1:n}) \nn \\[1mm]
&\hspace{7mm}= \rmv f(\V{x}_{0}) \ist \prod^{n}_{n'=1} \ist p\big(\V{a}_{n'}, M_{n'} \big|\V{x}_{n'} \big) \ist f\big(\V{x}_{n'} \big|\V{x}_{n'-1} \big). \label{eq:jointPrior} \\[-1mm]
\nn
\end{align}

\section{Problem Formulation and Estimation}
\label{sec:probFormEstimation}

At time $n$, our goal is to estimate the source state $\V{x}_{n}$ from the total observation vector $\V{z}_{1:n}$. For estimating $\V{x}_n$, we will develop an approximate calculation of the minimum mean-square error (MMSE) estimator \cite{Kay:B93}
\begin{equation}
\hat{\V{x}}^\text{MMSE}_{n} \,\triangleq\rmv \int\rmv \V{x}_n \ist f(\V{x}_n |\V{z}_{1:n}) \ist \mathrm{d}\V{x}_n \,.
\label{eq:mmse}
\end{equation}
This estimator involves the posterior PDF $f(\V{x}_n | \V{z}_{1:n})$ which is a marginal of the joint posterior PDF $f(\V{x}_{1:n}, \V{a}_{1:n}$ $| \V{z}_{1:n} )$ and involves all the source states, all the association variables, and all the observations, at all times up to the current time $n$.

In the following derivation of the factorization of $f(\V{x}_{1:n}, \V{a}_{1:n} | \V{z}_{1:n} )$, the observations $\V{z}_{1:n}$ are considered observed and thus fixed, 
and consequently the numbers of observations $\V{m}_{1:n}$ are fixed as well. Then, using Bayes' rule and the fact that $\V{z}_{1:n}$ implies $\V{m}_{1:n}$, we \vspace{1mm} obtain
\begin{align*}
&f(\V{x}_{1:n}, \V{a}_{1:n} | \V{z}_{1:n} ) \nn\\[1.5mm]
&\hspace{5mm}=\, f(\V{x}_{1:n}, \V{a}_{1:n}, \V{m}_{1:n}| \V{z}_{1:n} )\nn\\[1.5mm]
&\hspace{5mm}\propto\,  f( \V{z}_{1:n} | \V{x}_{1:n}, \V{a}_{1:n}, \V{m}_{1:n} ) \ist f(\V{x}_{1:n}, \V{a}_{1:n}, \V{m}_{1:n}).\\[-1mm]
\end{align*}
Inserting \eqref{eq:jointPrior} for $f(\V{x}_{1:n}, \V{a}_{1:n}, \V{m}_{1:n})$ and \eqref{eq:allTimeslhf} for $ f( \V{z}_{1:n} | \V{x}_{1:n},$ $\V{a}_{1:n}, \V{m}_{1:n} )$, then yields the final factorization
\begin{align}
f(\V{x}_{1:n}, \V{a}_{1:n} | \V{z}_{1:n} ) &\propto f( \V{x}_{0}) \prod^{n}_{n'= 1} \rmv \Psi\big(\V{a}_{n}\big) f( \V{x}_{n'} | \V{x}_{n'\rmv-1}) \nn \\[1mm]
&\hspace{10.3mm}\times  \prod^{K}_{k= 1} \hspace{.5mm} r_k\big( \V{x}_{n'}, a_{k,n'}; \V{z}_{n'} \big) \label{eq:factorOverall} \\[-4.5mm]
\nn \\[-10mm]
\nn
\end{align}
with\vspace{1mm} 
\begin{align}
r_k\big( \V{x}_{n}, a_{k,n}; \V{z}_{n}  \big) \triangleq\ist v\big( \V{x}_{n}, a_{k,n}; M_{n} \big) \ist h_k\big( \V{x}_{n}, a_{k,n}; \V{z}_{n} \big) \nn\\[-5mm]
\nn
\end{align}
and $\Psi\big(\V{a}_{n}\big) \triangleq | \Set{D}_{\V{a}_n} | ! \hspace{.7mm} \psi\big(\V{a}_{n}\big) $. The factor graph representing this factorization of the joint posterior PDF is depicted for one time step in Fig.\ \ref{fig:overallFG}.  By applying the sum-product algorithm (SPA) \cite{KscFreLoe:01} on this factor graph, accurate approximations \vspace{0mm} (``beliefs'') $\tilde{f}(\V{x}_n | \V{z}_{1:n})$ of the marginal posterior \vspace{0mm} PDFs $ f(\V{x}_n |\V{z}_{1:n})$, used for estimation in \eqref{eq:mmse}, are obtained in an efficient way.

\begin{figure}[t!]
\centering
\mbox{\hspace{-1.5mm}\begin{minipage}[H!]{0.24\textwidth}
\centering
\includegraphics[scale=.19]{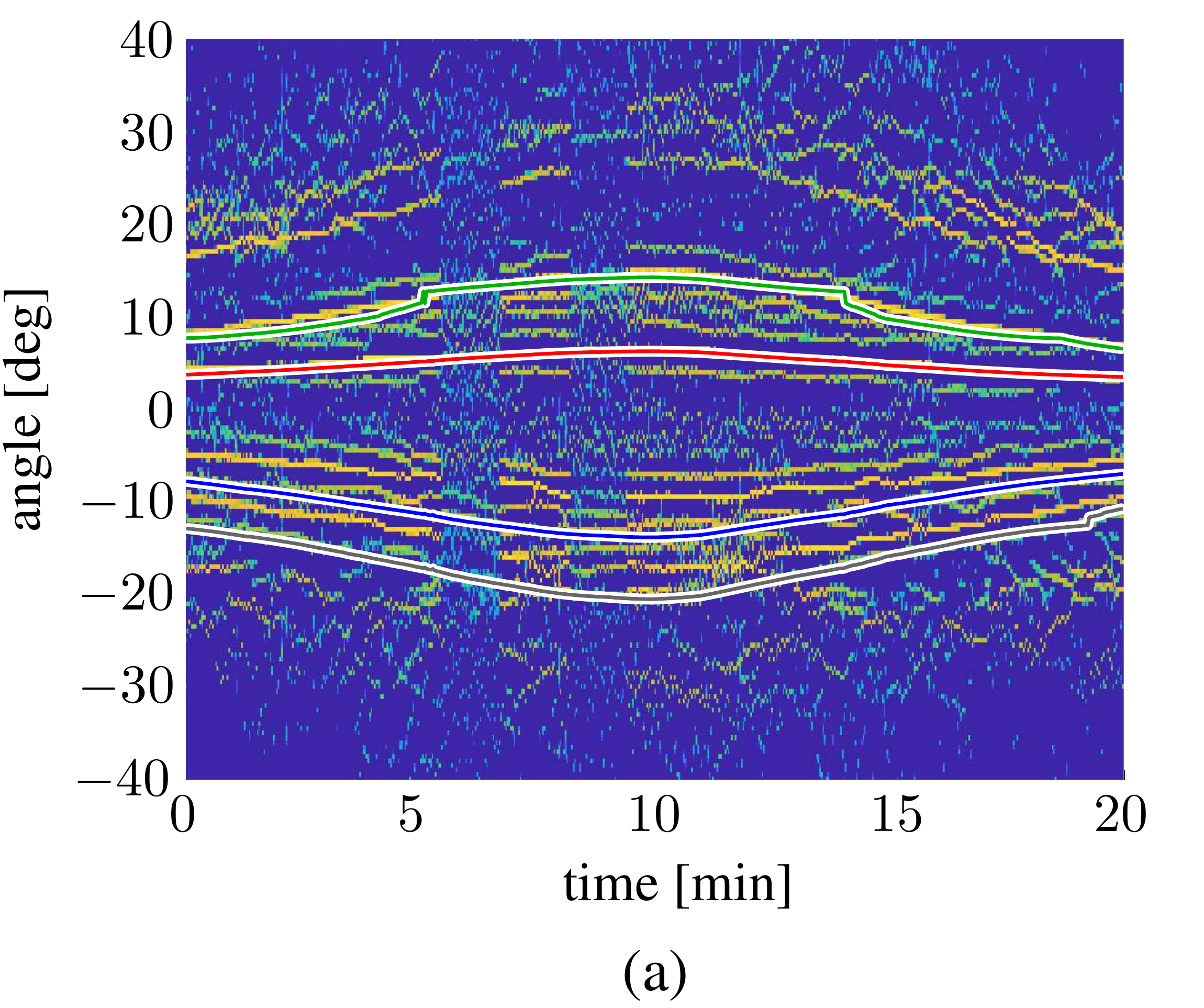}
\end{minipage}
\begin{minipage}[H!]{0.24\textwidth}
\centering
\hspace*{-1mm}\includegraphics[scale=.19]{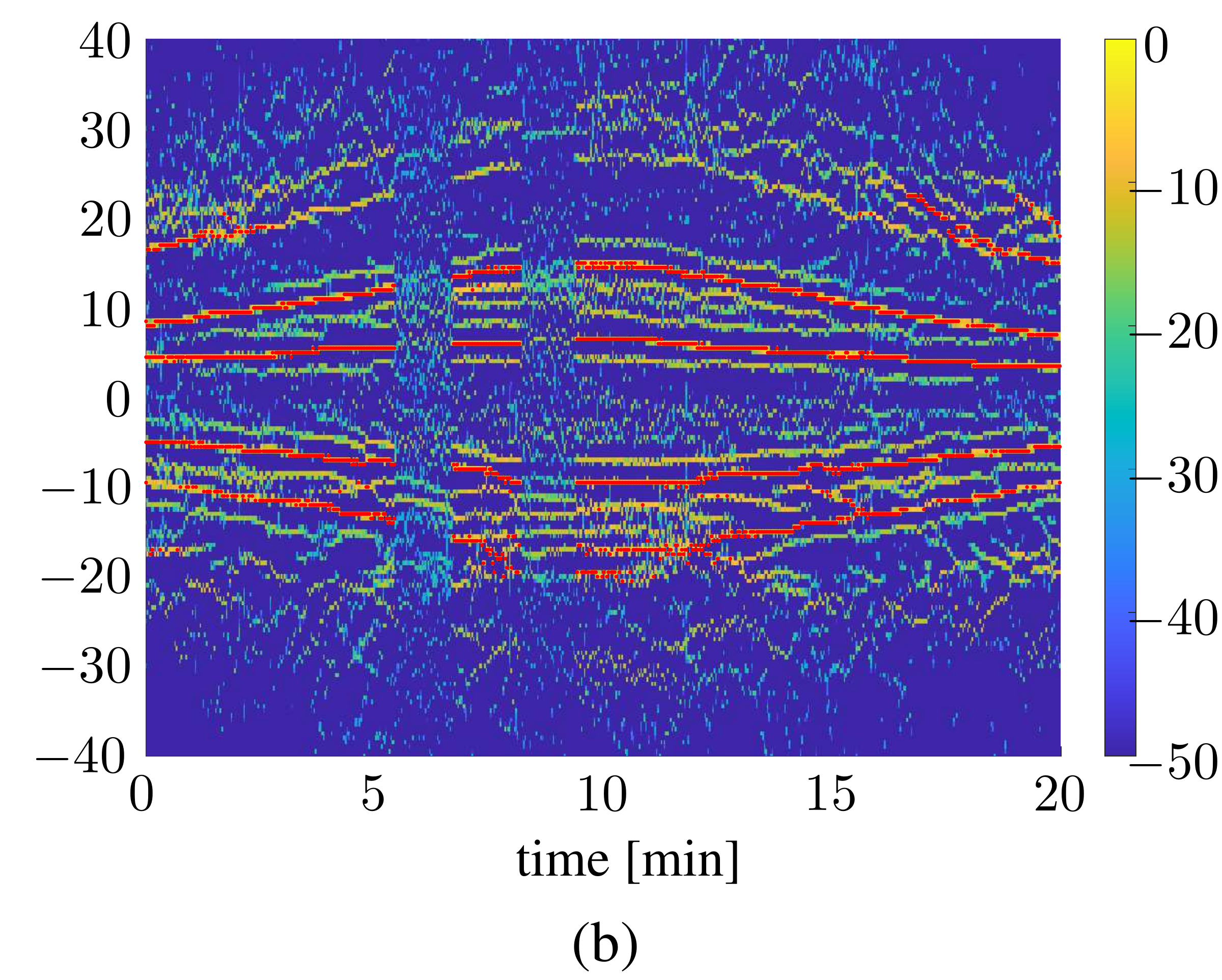}
\end{minipage}}
\renewcommand{\baselinestretch}{1.05}\small\normalsize
\vspace{0mm}
\caption{DOA measurements obtained by multi-frequency sparse Bayesian learning versus time and corresponding modeled DOAs. In (a) the expected DOAs related to the four dominate propagation paths are overlaid. These DOAs are obtained by using a source range calculate from the global positioning system (GPS) system of the R/V Sproul, a reference depth of $60$ m, and by applying the considered propagation model, respectively. In Fig.~\ref{fig:doaPlots}(b), the DOA estimates used as observations by the proposed method are highlighted in red. } 
\label{fig:doaPlots}
\vspace{-3mm}
\end{figure}

In order to obtain accurate nonlinear estimation by means of the SPA, we represent messages related to continuous state variables by random samples or particles\cite{AruMasGorCla:02,MeyBraWilHla:J17}. Thus, our algorithm  produces a particle representation $\big\{ \big( \V{x}_{n}^{(j)}\rmv\rmv, \omega_{n}^{(j)} \big) \big\}_{j=1}^{J}\rmv$ of the posterior PDF $ f(\V{x}_n |\V{z}_{1:n})$ at each time step $n$, i.e., a particle-based approximation of the posterior $f(\V{x}_n |\V{z}_{1:n})$ is given by $\tilde{f}(\V{x}_n |\V{z}_{1:n}) = \sum^{J}_{j = 1} \rmv \omega^{(j)}_n \ist \delta(\V{x}_n - \V{x}^{(j)}_n)$.  An \vspace{.2mm} approximation of the estimate $\hat{\V{x}}^{\text{MMSE}}_{n}$ in \eqref{eq:mmse} is then obtained from the respective particle representation \vspace{-1mm} as
\begin{equation}
\hat{\V{x}}_{n} \ist=\ist \sum_{j=1}^{J} \omega_{n}^{(j)} \V{x}_{n}^{(j)} \,. \nn
\vspace{-.5mm}
\end{equation}
For the considered estimation problem, the particle-based SPA is asymptotically optimum. This means that it can provide an approximation of the MMSE estimate in \eqref{eq:mmse} that can be made arbitrarily good by choosing $J$ sufficiently large \cite{AruMasGorCla:02}. For the processing of the \textit{SWellEx-96} data we set \vspace{2mm} $J = 10^4$.

\section{Ray Tracing}
\label{sec:impAspects}

At each time step $n$, the proposed particle-based algorithm evaluates the nonlinear DOA function $g_k(\V{p}_n)$ for each propagation path $k \in \{1,\dots,K\}$ a total of $J$ times. DOAs are obtained by using BELLHOP\cite{Por:I20} to calculate the eigenrays for the $J$ source positions based on a SSP along the water column and the bathymetry information.

To reduce the runtime of the proposed algorithm, we precompute a 3-D matrix $\M{Z} \in \mathbb{R}^{N_r \times N_d \times K}$ of DOAs that is used to interpolate the DOAs for each \vspace{-.3mm} propagation path $k \in \{1,\dots,K\}$ and each \vspace{-.2mm} source position $\V{p}^{(j)}_n$, $j \in \{1,\dots,J\}$  during runtime.

The matrix $\M{Z}$ is obtained by performing the following \vspace{-.7mm} steps:
\begin{enumerate}
\item we define a regular $N_r \times N_d$ grid of source positions $\V{p}^{(i_r,i_d)}_n \rmv \rmv$, $i_r \in \{1,\dots,N_r\}$, $i_d \in \{1,\dots,N_d\}$ that covers the entire region of  \vspace{-.7mm}interest; 
\item we use BELLHOP to calculate the eigenrays related to each source position on the grid $\V{p}^{(i_r,i_d)}_n \rmv \rmv$, $i_r \in \{1,\dots,N_r\}$, $i_d \in \{1,\dots,N_d\}$  and select the $K$ eigenrays that are expected to result in propagation  \vspace{-.7mm}paths for acoustic signals; and
\item we compute a 3-D matrix $\M{Z} \in \mathbb{R}^{N_r \times N_d \times K}$ that consists of the DOAs of the $K$ eigenrays at each of the $N_r \times N_d$ grid points; if eigenray $k \in \{1, \dots, K\}$ is geometrically impossible at a specific grid point, we set the corresponding element of matrix $\M{Z}$ to $-\infty$ and the corresponding detection probability $d_k\big(\V{p}^{(i_r,i_d)}_n\big)$ to zero.
\end{enumerate}

For the processing of the \textit{SWellEx-96} data we choose $N_r = 2400$ and $N_d = 165$, which correspond to a resolution of $1 \hspace{.5mm}$m.

\section{Results}
\label{sec:results}

We validate the proposed method  and compare it with matched field processing by using experimental data from a complex multi-path shallow-water environment \cite{DSpMurHodBooSch:J99,BooAbaSchHod:J00}. 

\subsection{Setup of Experiment and DOA Estimation}
Acoustic data sampled at $1500$ Hz was recorded by a 64-element vertical linear array with a uniform inter-sensor spacing of $1.875$ m that spanned water depths $94.125$-$212.25$ m. The surface ship R/V Sproul traveled with a radial speed of $2.5$ m/s towards the VLA with closest point of approach (CPA) at approximately 1 km. The ship towed an acoustic source at a depth of roughly $60$ m that projects an acoustic signal that includes 13 tones at frequencies $\{ 49, 64, 79, 94, 112,130,148,166,201,235,283,338,388\}$ Hz. The source track is shown in Fig.~\ref{fig:rayTracing}(a).

We process all 13 tones for the 20 minute interval indicated in Fig.~\ref{fig:rayTracing}(a). In particular, we perform beamforming by means of multifrequency sparse Bayesian learning (SBL)\cite{NanGemGerHodMec:J19} as follows. First, the data are split into $1758$ snapshots that consist of 2048 samples (1.37 s) and have 50 \% overlap. For each snapshot a rectangular window is applied and a fast Fourier transform (FFT) is performed.  The FFT length of 2048 samples corresponds to a bin width of $0.73$ Hz.  To accommodate Doppler shifts, for each of the 13 tones, we search the $\pm$1 FFT bins adjacent to the expected FFT bin and extract the FFT value with the maximum power. Next we perform multifrequency SBL for all tones and $L=3$ consecutive snapshots. This results in $586$ SBL solutions. Finally, we extract the DOA estimates from each SBL solution by finding (i) the 4 highest peaks and (ii) any additional peak that is within $35 \%$ of the highest peak. These DOA estimates are used as observations for the proposed method.

The SBL $586$ solutions are shown in Fig.~\ref{fig:doaPlots}.  In Fig.~\ref{fig:doaPlots}(a) and Fig.~\ref{fig:doaPlots}(b) expected DOAs related to the four dominate propagation paths are overlaid. These DOAs are obtained by using a source range calculate from the GPS system of the R/V Sproul, a reference depth of $60$ m, and by applying the considered propagation model. In Fig.~\ref{fig:doaPlots}(b), the DOA estimates used as observations by the proposed method are highlighted in red. Note that around time 7 min and 11 min for an interval of approximately 74 s, the transmission of the 13 tones was interrupted. The 72 data segments corresponding to these time intervals were discarded, i.e., no DOA estimates were extracted. The remaining 514 data segments define the number of time steps of the proposed method.

\begin{figure}[t!]
\centering
\begin{minipage}[H!]{0.6\textwidth}
\centering
\hspace*{-15mm}\includegraphics[scale=.23]{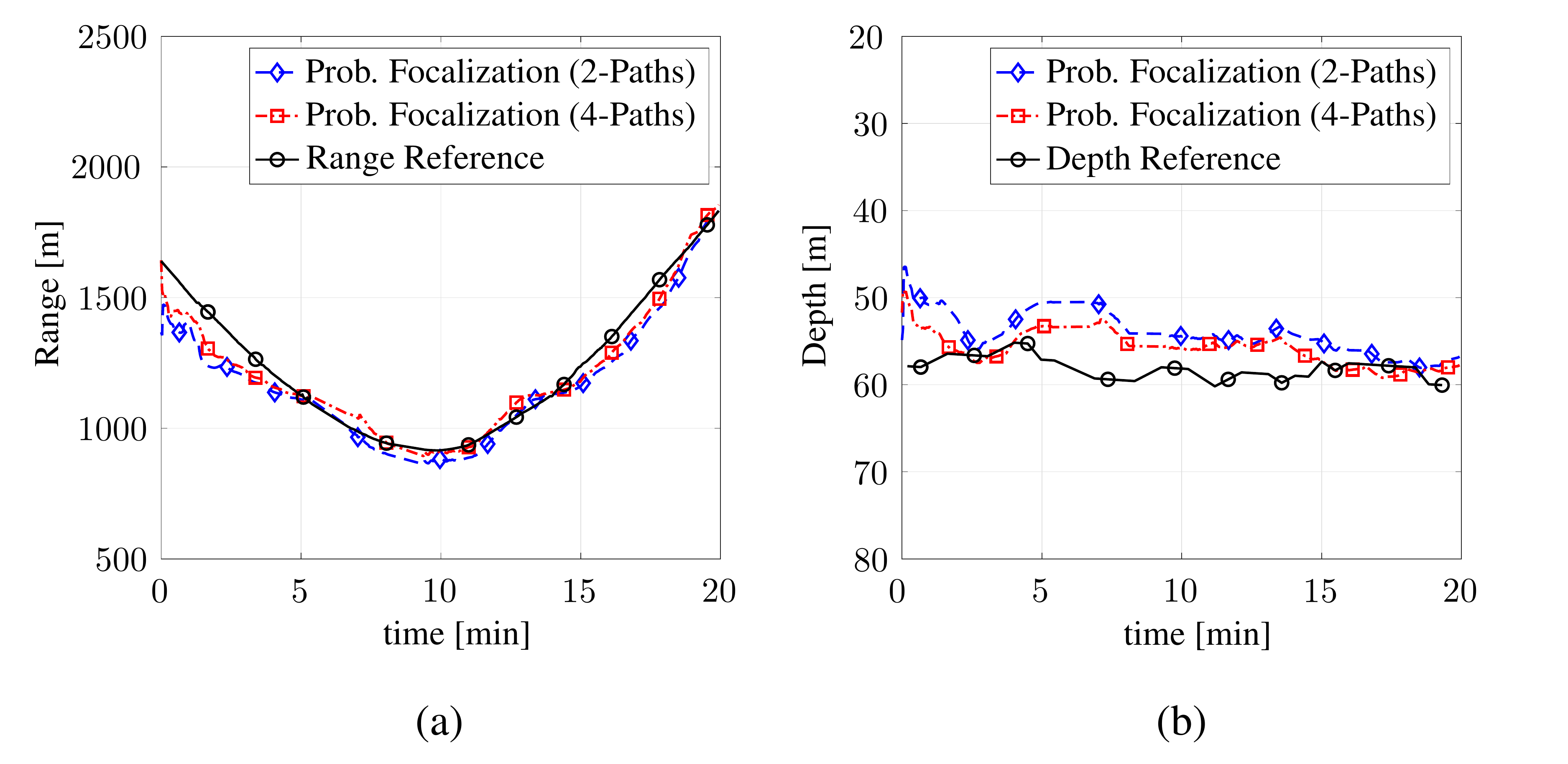}
\vspace*{1mm}
\end{minipage}
\vspace{-3mm}
\begin{minipage}[H!]{0.6\textwidth}
\centering
\hspace*{-20mm}\includegraphics[scale=.23]{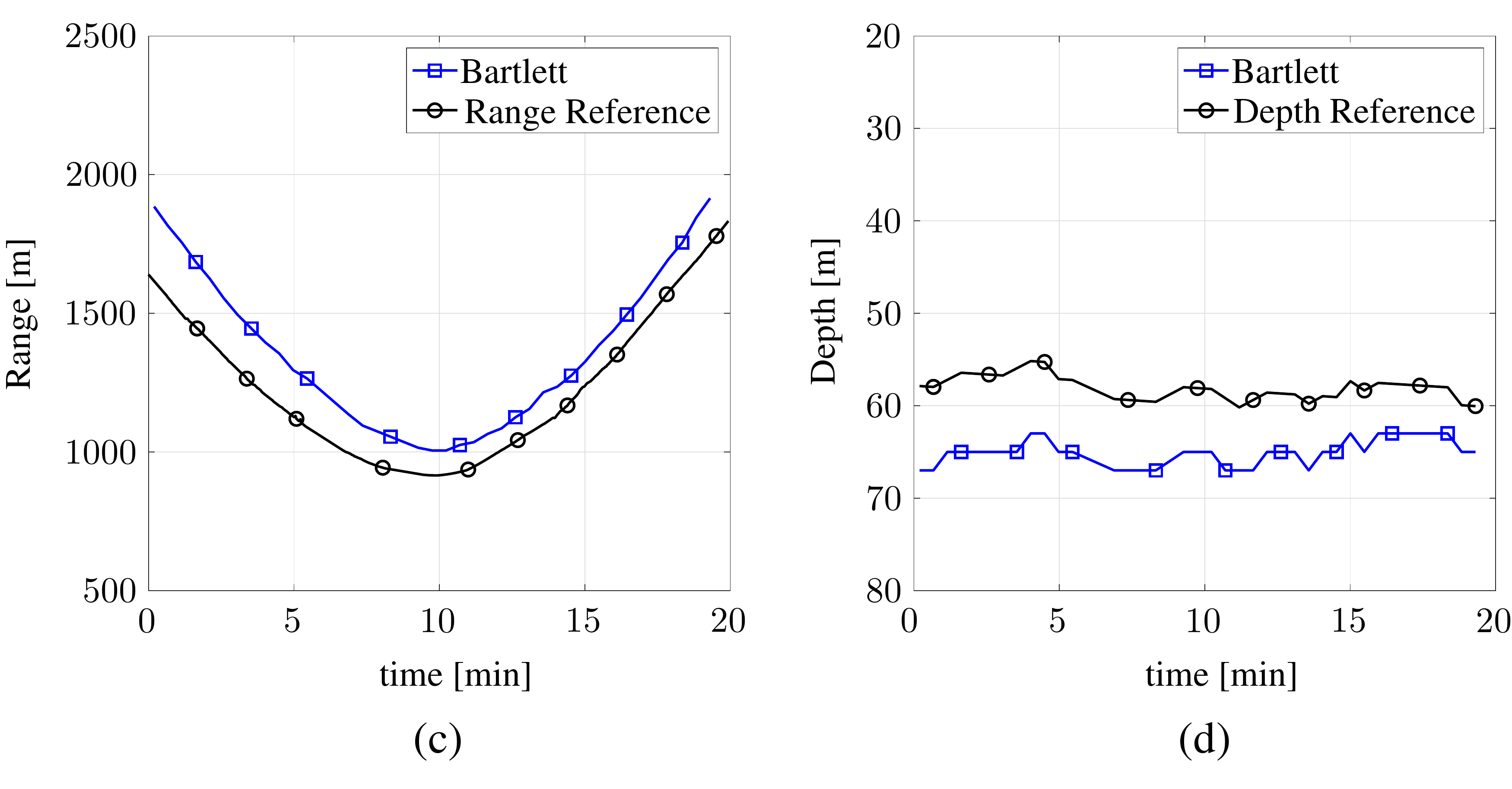}
\end{minipage}
\vspace*{0mm}
\caption{Shallow water source localization results for the considered dataset from the \emph{SwellEx-96} experiment. Probabilistic focalization results are shown in (a) \& (b). MFP results are shown in (c) \& (d). Compared to MFP, the proposed method is less sensitive to inaccurate bathymetry information.} 
\label{fig:results}
\vspace{-3.7mm}
\end{figure}

\subsection{System Parameters}

We consider a region of interest of $[100\ist\text{m}, \ist 2500\ist\text{m} ]  \times [10\ist\text{m}, \ist 175\ist\text{m}]$ in range and depth. Source motion is modeled by assuming a nearly constant-velocity model\cite{ShaKirLi:B02} in range and a nearly constant-location model in \vspace{-0.5mm} depth, i.e., 
\begin{equation}
\V{x}_{n} =  { \begin{pmatrix}
   1 & 0 & T_n  \\
   0 & 1 & 0  \\
   0  & 0  & 1 
  \end{pmatrix} } \ist\V{x}_{n-1} + { \begin{pmatrix}
   \ist\frac{T_n^2}{2} & 0 \\
   0  & T_n  \\
   T_n & 0 
  \end{pmatrix} } \V{u}_{n}
  \label{eq:stateTransition}
  \vspace{0mm}
  \end{equation}
where $\V{u}_{n}$ is the driving noise and $T_n$ is the length of the time step. If there is no data dropout, we have $T_n = 2.048$s. If data segments are missing, $T_n$ is increased accordingly.

The driving noise $\V{u}_{n} \rmv\sim \Set{N}(\V{0},\M{\Sigma}_u )$ with $\M{\Sigma}_u  \!= \mathrm{diag} \big\{0.05 \ist\ist \text{m}^2/\text{s}^4 \hspace{2.5mm} 0.1 \ist\ist \text{m}^2/\text{s}^2 \big\}$ is an independent and identically distributed (iid) sequence of 2D Gaussian random vectors. Note that \eqref{eq:stateTransition} fully defines the state transition function $f(\V{x}_{n}|\V{x}_{n-1})$ discussed in Section \ref{sec:vec_description}.

We process the DOA observations with the proposed method by considering the $K=4$ propagation paths shown it Fig.~\ref{fig:rayTracing}. Furthermore, we also process the DOA observations by only considering the $K=2$ propagation paths that are not affected by the seabed (DP and SB). Our model uses the SSP measured on the day of the event and shown in Fig.~\ref{fig:rayTracing}.

The observation noise standard deviations (cf.~\eqref{eq:measModel}) are set to $\sigma_{k} = 0.5^{\circ}$ for $k \rmv\in\rmv \{1,2\}$ and $\sigma_{k} = 2^{\circ}$ for $k \rmv\in\rmv \{3,4\}$. The larger standard deviation for the propagation paths that involve bottom bounces is motivated by the fact that constant bathymetry used by the model is inaccurate. The detection probabilities are set to $d_k(\V{p}_{n}) \rmv=\rmv 0.9$ if propagation path $k$ is geometrically possible and set to zero otherwise. The false alarm PDF $f_{\text{FA}}\big( z_{m,n} \big)$ is uniform on $[-90^{\circ},90^{\circ})$. The mean number of false alarms is $\mu_{\text{FA}} \!\rmv=\! 2$ for $K \rmv=\rmv 4$ and $\mu_{\text{FA}} \!\rmv=\! 4$ for $K \rmv=\rmv 2$. At time $n$, the prior distribution has the form $f( \V{x}_{0}) = f( \V{p}_{0}) f( v_{0})$, where $f( \V{p}_{0})$ is uniform on $[100\ist\text{m}, \ist 2500\ist\text{m} ]  \times [10\ist\text{m}, \ist 175\ist\text{m}]$ and $f( v_{0})$ is zero-mean Gaussian with standard deviation $5$ m/s. Our implementation of the proposed method used $J \rmv=\rmv 10^4$ particles.

\subsection{Performance Comparison}

As a reference method, we consider MFP by means of the Bartlett processor. Both methods use the SSP, a constant bathymetry of $216.5$ m, as well as the geo-acoustic parameters discussed in Ref.~\cite{GemNanGerHod:J17}. As a reference for the true range of the source, we use the heading and GPS position of the R/V Sproul and assume that the source is towed $100$ m behind the vessel. As a reference for true source depth, we use the MFP solution compensated for bathymetry mismatch based on Eq.~(10) in Ref.~\cite{DSpMurHodBooSch:J99} and true seabed depths at the source locations. 

Fig.~\ref{fig:results} shows the shallow water source localization results for the considered  $20$ min of data. Range and depth estimation results for probabilistic focalization are shown in Fig.~\ref{fig:results} (a) \& (b).  While only $K \rmv=\rmv2$ propagation paths are needed for localization, using $K \rmv=\rmv4$ instead of $K \rmv=\rmv2$ propagation paths can slightly increase the range and depth estimation accuracy of the proposed method.

MFP results are shown in Fig.~\ref{fig:results} (c) \& (d). The range and depth bias of the MFP solution, also known as the mirage effect\cite{DSpMurHodBooSch:J99}, is related to the fact that the assumed constant bathymetry is not accurate. In the considered scenario, the proposed method can outperform MFP despite relying on less environmental information. In particular, compared to MFP it is less sensitive to inaccurate bathymetry information. This is enabled by the proposed statistical model which makes it possible to assign different observation noise uncertainties to different propagation\vspace{0.5mm} paths.

\section{Conclusion}
\label{sec:conclusion}

We introduced a probabilistic focalization approach for the localization and tracking of an acoustic source in shallow water. Our method probabilistically associates observed DOAs to modeled DOAs and jointly estimates the time-varying source location. We demonstrated performance advantages compared to MFP using data collected during the  \emph{SWellEx-96}  experiment. Notably, despite relying on less environmental information, the proposed method can outperform MFP in terms of range and depth estimation\vspace{1.5mm} accuracy.

\section{Acknowledgement}
This research was supported by the Office of Naval Research under Grants N00014-21-1-2267 and N00014-21-WX-0-1634. We thank Prof.~William S.~Hodgkiss and Dr.~Peter Gerstoft for helpful discussions. 
\vspace{2mm}

\renewcommand{\baselinestretch}{1}
\selectfont
\bibliographystyle{IEEEtran}
\bibliography{IEEEabrv,Books,Papers,Temp}

\begin{thebibliography}{10}
\providecommand{\url}[1]{#1}
\csname url@samestyle\endcsname
\providecommand{\newblock}{\relax}
\providecommand{\bibinfo}[2]{#2}
\providecommand{\BIBentrySTDinterwordspacing}{\spaceskip=0pt\relax}
\providecommand{\BIBentryALTinterwordstretchfactor}{4}
\providecommand{\BIBentryALTinterwordspacing}{\spaceskip=\fontdimen2\font plus
\BIBentryALTinterwordstretchfactor\fontdimen3\font minus
  \fontdimen4\font\relax}
\providecommand{\BIBforeignlanguage}[2]{{%
\expandafter\ifx\csname l@#1\endcsname\relax
\typeout{** WARNING: IEEEtran.bst: No hyphenation pattern has been}%
\typeout{** loaded for the language `#1'. Using the pattern for}%
\typeout{** the default language instead.}%
\else
\language=\csname l@#1\endcsname
\fi
#2}}
\providecommand{\BIBdecl}{\relax}
\BIBdecl

\bibitem{KupHodHeeAkaFerJac:J98}
W.~A. Kuperman, W.~S. Hodgkiss, H.-C. Song, T.~Akal, C.~Ferla, and D.~R.
  Jackson, ``{Phase conjugation in the ocean: Experimental demonstration of an
  acoustic time-reversal mirror},'' \emph{J. Acoust. Soc. Am.}, vol. 103,
  no.~1, pp. 25--40, 1998.

\bibitem{Hin:J73}
M.~J. Hinich, ``{Maximum-likelihood signal processing for a vertical array},''
  \emph{J. Acoust. Soc. Am.}, vol.~54, no.~2, pp. 499--503, 1973.

\bibitem{Buc:J76}
H.~P. Bucker, ``{Use of calculated sound fields and matched-field detection to
  locate sound sources in shallow water},'' \emph{J. Acoust. Soc. Am.},
  vol.~59, no.~2, pp. 368--373, 1976.

\bibitem{BagKupMik:J93}
A.~B. {Baggeroer}, W.~A. {Kuperman}, and P.~N. {Mikhalevsky}, ``An overview of
  matched field methods in ocean acoustics,'' \emph{IEEE Journal of Oceanic
  Engineering}, vol.~18, no.~4, pp. 401--424, 1993.

\bibitem{ColFiaKup:J95}
M.~D. Collins, L.~T. Fialkowski, W.~A. Kuperman, and J.~S. Perkins, ``{The
  multivalued Bartlett processor and source tracking},'' \emph{J. Acoust. Soc.
  Am.}, vol.~97, no.~1, pp. 235--241, 1995.

\bibitem{ThoKupDsp:J00}
A.~M. Thode, W.~A. Kuperman, G.~L. D'Spain, and W.~S. Hodgkiss, ``{Localization
  using Bartlett matched-field processor sidelobes},'' \emph{J. Acoust. Soc.
  Am.}, vol. 107, no.~1, pp. 278--286, 2000.

\bibitem{ByuVerSab:J17}
S.-H. Byun, C.~M.~A. Verlinden, and K.~G. Sabra, ``{Blind deconvolution of
  shipping sources in an ocean waveguide},'' \emph{J. Acoust. Soc. Am.}, vol.
  141, no.~2, pp. 797--807, 2017.

\bibitem{GemNanGerHod:J17}
K.~L. Gemba, S.~Nannuru, P.~Gerstoft, and W.~S. Hodgkiss, ``{Multi-frequency
  sparse Bayesian learning for robust matched field processing},'' \emph{J.
  Acoust. Soc. Am.}, vol. 141, no.~5, pp. 3411--3420, 2017.

\bibitem{GemHodGer:J17}
K.~L. Gemba, W.~S. Hodgkiss, and P.~Gerstoft, ``Adaptive and compressive
  matched field processing,'' \emph{J. Acoust. Soc. Am.}, vol. 141, no.~1, pp.
  92--103, 2017.

\bibitem{OrrNicPer:J20}
G.~J. Orris, M.~Nicholas, and J.~S. Perkins, ``{The matched-phase coherent
  multi-frequency matched-field processor},'' \emph{J. Acoust. Soc. Am.}, vol.
  107, no.~5, pp. 2563--2575, 2000.

\bibitem{Kro:J92}
J.~L. Krolik, ``{Matched-field minimum variance beamforming in a random ocean
  channel},'' \emph{J. Acoust. Soc. Am.}, vol.~92, no.~3, pp. 1408--1419, 1992.

\bibitem{ByuHunGemSonKup:J20}
G.~Byun, F.~H. Akins, K.~L. Gemba, H.~C. Song, and W.~A. Kuperman, ``Multiple
  constraint matched field processing tolerant to array tilt mismatch,''
  \emph{J. Acoust. Soc. Am.}, vol. 147, no.~2, pp. 1231--1238, 2020.

\bibitem{MeyBraWilHla:J17}
F.~Meyer, P.~Braca, P.~Willett, and F.~Hlawatsch, ``{A scalable algorithm for
  tracking an unknown number of targets using multiple sensors},'' \emph{{IEEE}
  Trans. Signal Process.}, vol.~65, no.~13, pp. 3478--3493, 2017.

\bibitem{MeyKroWilLauHlaBraWin:J18}
F.~Meyer, T.~Kropfreiter, J.~L. Williams, R.~A. Lau, F.~Hlawatsch, P.~Braca,
  and M.~Z. Win, ``Message passing algorithms for scalable multitarget
  tracking,'' \emph{Proc. {IEEE}}, vol. 106, no.~2, pp. 221--259, 2018.

\bibitem{LeiMeyHlaWitTufWin:J19}
E.~Leitinger, F.~Meyer, F.~Hlawatsch, K.~Witrisal, F.~Tufvesson, and M.~Z. Win,
  ``{A Belief Propagation Algorithm for Multipath-Based SLAM},'' \emph{{IEEE}
  Trans. Wireless Commun.}, vol.~18, no.~11, pp. 5613--5629, 2019.

\bibitem{MenMeyBauWin:J19}
R.~Mendrzik, F.~Meyer, G.~Bauch, and M.~Z. Win, ``Enabling situational
  awareness in millimeter wave massive {MIMO} systems,'' \emph{{IEEE} J. Sel.
  Topics Signal Process.}, vol.~13, no.~5, pp. 1196--1211, 2019.

\bibitem{MeyWin:J20}
F.~{Meyer} and M.~Z. {Win}, ``Scalable data association for extended object
  tracking,'' \emph{IEEE Trans. Signal Inf. Process. Netw.}, vol.~6, pp.
  491--507, May 2020.

\bibitem{MeyWil:J21}
F.~Meyer and J.~L. Williams, ``Scalable detection and tracking of geometric
  extended objects,'' 2021, arXiv:2103.11279.

\bibitem{Col:J91}
M.~D. Collins and W.~A. Kuperman, ``{Focalization: Environmental focusing and
  source localization},'' \emph{J. Acoust. Soc. Am.}, vol.~90, no.~3, pp.
  1410--1422, 1991.

\bibitem{NanGemGerHodMec:J19}
S.~Nannuru, K.~L. Gemba, P.~Gerstoft, W.~S. Hodgkiss, and C.~F.
  Mecklenbr{\"{a}}uker, ``{Sparse Bayesian learning with multiple
  dictionaries},'' \emph{Signal Process.}, vol. 159, pp. 159--170, 2019.

\bibitem{MeyParGer:C20}
F.~{Meyer}, Y.~{Park}, and P.~{Gerstoft}, ``Variational {Bayesian} estimation
  of time-varying {DOAs},'' in \emph{Proc. FUSION-2020}, Pretoria, South
  Africa, 2020.

\bibitem{BarWilTia:B11}
Y.~Bar-Shalom, P.~K. Willett, and X.~Tian, \emph{{Tracking and Data Fusion: A
  Handbook of Algorithms}}.\hskip 1em plus 0.5em minus 0.4em\relax Storrs, CT:
  Yaakov Bar-Shalom, 2011.

\bibitem{FinKupPor:B11}
F.~B. Jensen, W.~A. Kuperman, M.~B. Porter, and H.~Schmidt, \emph{Computational
  Ocean Acoustics}, 2nd~ed.\hskip 1em plus 0.5em minus 0.4em\relax New York,
  NY: Springer, 2011.

\bibitem{Por:I20}
M.~B. Porter~et al., ``{The Acoustics Toolbox},'' Online, 2020, available:
  {http://oalib.hlsresearch.com/AcousticsToolbox/}.

\bibitem{Kay:B93}
S.~M. Kay, \emph{{Fundamentals of Statistical Signal Processing: Estimation
  Theory}}.\hskip 1em plus 0.5em minus 0.4em\relax Upper Saddle River, NJ:
  Prentice-Hall, 1993.

\bibitem{KscFreLoe:01}
F.~R. Kschischang, B.~J. Frey, and H.-A. Loeliger, ``Factor graphs and the
  sum-product algorithm,'' \emph{{IEEE} Trans. Inf. Theory}, vol.~47, no.~2,
  pp. 498--519, Feb. 2001.

\bibitem{AruMasGorCla:02}
M.~S. Arulampalam, S.~Maskell, N.~Gordon, and T.~Clapp, ``{A tutorial on
  particle filters for online nonlinear/non-Gaussian {B}ayesian tracking},''
  \emph{{IEEE} Trans. Signal Process.}, vol.~50, no.~2, pp. 174--188, 2002.

\bibitem{DSpMurHodBooSch:J99}
G.~L. D'Spain, J.~J. Murray, W.~S. Hodgkiss, N.~O. Booth, and P.~W. Schey,
  ``{Mirages in shallow water matched field processing},'' \emph{J. Acoust.
  Soc. Am.}, vol. 105, no.~6, pp. 3245--3265, 1999.

\bibitem{BooAbaSchHod:J00}
N.~O. Booth, A.~T. Abawi, P.~W. Schey, and W.~S. Hodgkiss, ``{Detectability of
  low-level broad-band signals using adaptive matched-field processing with
  vertical aperture arrays},'' \emph{{IEEE} J. Ocean. Eng.}, vol.~25, no.~3,
  pp. 296--313, 2000.

\bibitem{ShaKirLi:B02}
Y.~Bar-Shalom, T.~Kirubarajan, and X.-R. Li, \emph{{Estimation with
  Applications to Tracking and Navigation}}.\hskip 1em plus 0.5em minus
  0.4em\relax New York, NY, USA: Wiley, 2002.

\end{thebibliography}

\end{document}